\documentstyle[preprint,revtex]{aps}
\begin{document}
\begin{title}
Electronic structure of superconducting ${\rm Ba_6C_{60}}$
\end{title}
\author{Steven C.\ Erwin}
\begin{instit}
Department of Physics, University of Pennsylvania, Philadelphia,
Pennsylvania 19104
\end{instit}
\author{Mark R.\ Pederson}
\begin{instit}
Complex Systems Theory Branch, Naval Research Laboratory,
Washington, D.C.\ 20375
\end{instit}
\begin{abstract}
We report the results of first-principles electronic-structure
calculations for superconducting ${\rm Ba_6C_{60}}$.  Unlike the
$A_3{\rm C}_{60}$ superconductors, this new compound shows strong
Ba--C hybridization in the valence and conduction regions,
mixed covalent/ionic bonding character, partial charge transfer,
{\it and insulating zero-gap band structure}.
\end{abstract}
Most of the known superconducting C$_{60}$ fullerides exhibit the
stoichiometry
$A_3{\rm C}_{60}$ or $A_2B{\rm C}_{60}$ ($A,B$=alkali), form fcc
lattices,
and are  known (or believed) to exhibit essentially complete charge
transfer
from alkali to C$_{60}$. Broadly speaking, all of these compounds
exhibit similar
electronic structure, in which the half-filled valence band (VB) is
formed
primarily from C orbitals of $p_z$ symmetry.
Superconductivity has
also been observed in
5:1 solid solutions of Ca in C$_{60}$ \cite{koretal92a}.  In
contrast to the
$A_3{\rm C}_{60}$ compounds,
theoretical studies of
hypothetical (stoichiometric) Ca$_n$C$_{60}$ indicate
substantial hybridization of C and Ca orbitals in the VB regime
\cite{saiosh92},
but the relevance of these results to the real (non-stoichiometric)
material is
difficult to gauge.

Recently, synthesis of a crystalline, stoichiometric,
Ba-intercalated fulleride superconductor, ${\rm Ba_6C_{60}}$, has been
reported,
with $T_{\rm c}$=7 K \cite{koretal92b}.
Naive electron
counting, assuming divalent Ba and rigid ${\rm K_6C_{60}}$-like
bandstructure,
would suggest this new compound to be a 0.2-eV gap insulator consisting
of very
highly charged C$^{-12}_{60}$ molecules. Here, we present
first-principles
electronic structure results for ${\rm Ba_6C_{60}}$ that
definitively predict
{\it strong Ba-C hybridization near the Fermi level,
incomplete charge transfer and,
surprisingly, a zero-gap insulating band structure.}

Like the alkali $A_6{\rm C}_{60}$ compounds, ${\rm Ba_6C_{60}}$
forms a
bcc Bravais lattice, with two-fold planar crosses of Ba ions
decorating the
faces of the conventional unit cell.  The aspect ratio of the
crosses is
identical to that in ${\rm K_6C_{60}}$ \cite{zhoetal91}, and the
lattice constant of
11.171 \AA~is 2\% smaller than for ${\rm K_6C_{60}}$. As with
${\rm K_6C_{60}}$, Rietveld refinement of x-ray powder
patterns
for ${\rm Ba_6C_{60}}$ indicate rotationally ordered
C$_{60}$ molecules.

We have studied the electronic structure and energetics of
crystalline ${\rm
Ba_6C_{60}}$ using the local-density approximation (LDA) to
density-functional theory.  The computational methods were identical
to
those used in previous studies of ${\rm K_6C_{60}}$ and ${\rm
K_3C_{60}}$ \cite{erwped91,erwpic91}, with a few modifications.  In
our
local-orbital method \cite{erwpedpic90}, basis functions are
expanded on a set
of fourteen Gaussian exponents contracted into four
$s$-type and three $p$-type
functions for C, and 21 exponents contracted into seven
$s$-type, five
$p$-type, and four $d$-type functions for Ba.  The smallest Gaussian
exponents
were 0.24 for C and 0.15 for Ba;  these basis sets give eigenvalue
spectra
converged to a tolerance of $\sim$0.02 eV for core, valence, and
low-lying
conduction bands.  The charge density and potential are completely
general and
without shape approximation, and were iterated to self-consistency
using the
$\Gamma$ point. All 696 electrons were explicitly included in the
calculation
and were treated on equal footing, without the need for
pseudopotentials. The
Ceperly-Alder exchange-correlation functional was used. Lowest-order
relativistic effects were accounted for perturbatively, as described
below.

The resulting band structure and density of states (DOS) are shown
in Fig.\
\ref{fig:bands}, plotted along the same high-symmetry lines as for
our previous
study of  ${\rm K_6C_{60}}$ ({\it cf.} Fig.\ 2 of Ref.\
\cite{erwped91}). Although the
correspondence between the two is clear, substantial differences
arise from
strong hybridization of C $2p$ and Ba $5d$ wave functions in the
valence and
conduction regions. Both the $t_{1u}$ band (filled in
${\rm K_6C_{60}}$) and the
$t_{1g}$ band (empty in ${\rm K_6C_{60}}$) are filled in
${\rm Ba_6C_{60}}$.
Surprisingly, the electronic structure is again insulating, since
there are no
partially filled bands and the Fermi-level DOS is indeed zero. The
magnitude of
the gap is zero, to within the limits of accuracy of our
calculation.  This is
strikingly  at odds with the experimental finding of
superconductivity (and
therefore a metallic normal state) in ${\rm  Ba_6C_{60}}$.
Moreover, we note
that while the LDA is known generally to give semiconductor band
gaps smaller
than experiment, we are unaware of any instance for which the LDA
band
structure of a known metal is insulating. We have exhaustively
tested the
quality of our basis set:  for a variety of different sets of
Gaussian exponents,
the longest-ranged of which was 0.12 for C and 0.10 for Ba, the VB
and CB varied
by at most 0.02 eV, and by less than this near the Fermi level.

A enlarged view of the DOS near the Fermi level is shown in
Fig.~\ref{fig:pdos}, along with atom- and symmetry-projected
partial DOS.  States within 1 eV of $E_{\rm F}$ have approximately
75\% C 2$p$ character and 25\% Ba 5$d$ character,
with a very small admixture
of C $s$ character in the low-lying CB.  The
contribution from Ba $s$ and
$p$ basis functions to bands in this energy region is negligible.

We have also simulated relativistic corrections to the bands, by
appending to the
Hamiltonian a non-local operator diagonal in the atomic-orbital
Bloch
basis.  These atomic shifts were taken over directly from
differences between
relativistic and non-relativistic atomic eigenvalues for Ba (the
shifts for C are
negligible).  For Ba core states, the shifts are large and resulted
in purely
rigid shifts of the corresponding bands.  For the 5$d$ states, which
contribute to the VB and lower CB, we  used the degeneracy-weighted
average of the atomic
5$d_{3/2}$ and 5$d_{5/2}$ shifts \cite{erwpedunpub}, giving a single
shift of
$+$0.44 eV.    Since the  upper VB
and lower CB states  have comparable Ba 5$d$ admixture,
diagonalization
 of this new Hamiltonian resulted in  nearly rigid band shifts
in this energy region.
Indeed, within 1
eV of $E_{\rm F}$, the changes were at most 0.1 eV, and the zero-gap
insulating
character of the band structure was unchanged (the relativistically
corrected bands are shown in two of the panels of Fig.\
\ref{fig:bands}, and the partial DOS in Fig.~\ref{fig:pdos}
includes this correction).
Relativistic shifts for the Ba 6$s$ atomic orbitals
are quite small ($-$0.16 eV), and since these states
contribute only to bands
several eV above $E_{\rm F}$,
these shifts have a negligible effect on
the occupied eigenvalue spectrum and charge density.

We have not systematically attempted to include spin-orbit (SO)
effects here.
The atomic SO splitting of the Ba 5$d_{3/2}$ and 5$d_{5/2}$ states
is 0.11 eV
\cite{erwpedunpub}.  To a first approximation, one expects the
resultant band
splitting also to be  proportional to the amount of 5$d$  character,
and so we
estimate the SO splitting in both the upper VB and lower CB to  be
of order 0.05
eV or less.  Shifts of $\pm$0.05 eV to $E_{\rm F}$ would result in a
Fermi-level DOS,
$N(E_{\rm F})$, of order 1--2 states/eV-cell-spin.  This is 3--6 times
lower
than what we have previously calculated for ${\rm K_3C_{60}}$
\cite{erwpic91}.
If the pairing mechanism is the same in both materials and the
coupling
strengths are comparable, then such values for $N(E_{\rm F})$ are
far too
low to account for the measured $T_{\rm c}$ of 7 K.   This
suggests a number
of possible scenarios: (1)  If SO splitting is indeed responsible
for making the
bands metallic, then either the ${\rm Ba_6C_{60}}$
superconducting mechanism itself or the pairing strength (or both)
is quite
different from the standard fullerene superconductors.
(2) If the
one-electron
(mean-field) approximation of the LDA does not apply to
${\rm Ba_6C_{60}}$,
then it is conceivable that a full accounting of electron-electron
correlation leads to a metallic normal state. We
view this
as unlikely, since the standard result of treating electron correlation
more
accurately is to increase the band gap, or even to give an
insulating ground state
when the single-particle ground state is metallic.
(3) In principle, it is possible that a self-consistent
{\it spin-polarized} calculation would lead to a
metallic ground state.
Such magnetic instabilities generally have as prerequisite a large
value for $N(E_{\rm F})$ in the paramagnetic (spin-unpolarized)
state.  Since we find a paramagnetic solution with
$N(E_{\rm F})\approx 0$,
we view this scenario as unlikely.
(4) If we take seriously the zero-gap insulating state,
then the possibility also exists that BCS-like pairing occurs
not between states
on the Fermi surface, but rather between states whose
energies lie in some small interval $E_{\rm F}\pm\Delta E$. Clearly,
this picture would be favored by the existence of an
extremely small, or zero, gap.
Interestingly, this scenario also requires some minimum
``threshold'' coupling strength, below which pairing does not occur
(in contrast to pairing of states on the
Fermi surface, for which {\it any} non-zero coupling suffices).

We consider now the question of charge transfer from Ba to C$_{60}$.
In a
previous study of ${\rm K_6C_{60}}$, we used Mulliken population
analysis to
compute the charge associated with each atom in the cell, and found
essentially
complete charge transfer from K to C$_{60}$ \cite{erwped91}. For
${\rm Ba_6C_{60}}$, Mulliken analysis associates 55.57 electrons
with each Ba,
suggesting a loss of only 0.43 electrons per atom. We caution,
however, that
Mulliken charges may not be meaningful when there is large
wave-function
overlap, and so we turn to direct methods.  Integration of the total
charge
density within touching muffin-tin (MT) spheres, of radius $R$,
gives the
charges, $Q(R)$, listed in the third column of Table
\ref{table:charges}. Of
course, these spheres do not represent the entire cell volume, nor
are they the only plausible choice of integration volumes. Thus,
the total integrated charges are less meaningful than  the
difference
charges (relative to overlapping neutral atoms),
$\Delta Q(R)$, shown in the fourth column.
These values suggest that $\sim$0.7 electrons are transferred
from each Ba
atom, in
rough agreement with the Mulliken result.

A more detailed description of the spatial distribution of Ba charge
is given by
the function $\Delta Q(r)$ for values of $r$ less than the touching
MT radii.  For
Ba, this function is shown in Fig.\ \ref{fig:deltaq}, for
$R_{\rm C}\leq r\leq
R_{\rm Ba}$.  It is clear that most of the 0.6 electron is lost from
the region
1.6 \AA~$\leq r \leq 2.2$ \AA. Moreover, we find that the Ba $5d$
and $6s$
atomic expectation values of $r$ are 1.65 \AA~and 2.33
\AA,~respectively \cite{erwpedunpub}. These observations are
consistent with
a picture in which charge is lost from the Ba $6s$ state, and
donated partially to
Ba $5d$ states and partially to C $p$ states. This description is
also consistent
with our prediction of the mixed $5d/2p$ character of the
highest-lying occupied
bands.

Some insight into the differences between the K- and
Ba-intercalated
fulleride crystals can be gained by examining the breakdown of an
ionic model that
was found to successfully describe ${\rm K_6C_{60}}$. Since the details
of this model have been described elsewhere,\cite{richmondbb} we only
present a brief qualitative discussion here.
For a system of non-overlapping
entitities, the charge transfer can be predicted by properly
accounting for the
Madelung contributions and the density-functional-based electron
affinities (for the
C$_{60}$ molecule) and the ionization energies (for the metallic
dopants).
When
such a model is applied to ${\rm K_6C_{60}}$, full charge
transfer is expected
at the experimental lattice constant of 11.39 \AA. Furthermore,
since at
this lattice constant
the smallest C--K distance (3.2 \AA) is large compared to the sum of
the ionic radii
(2.1 \AA), corrections due to banding are expected to be small.
Results from this
ionic model
were in excellent agreement with the charge transfer
calculated self-consistently for ${\rm K_6C_{60}}$ \cite{erwped91}.
In contrast, when the same
ionic model is applied
to ${\rm Ba_6C_{60}}$, we find that the Ba atoms are
expected to lose 1.2 electrons,
roughly twice as much as calculated self-consistently
in the crystal. This
discrepancy can be understood by noting that a Ba atom in such
a charge state
would still have excess valence charge available for bonding with
the C atoms.
Further, for this charge state, the
Ba $5d$ states exhibit a maximum at a radius of approximately
1.5--1.6 \AA~and
exhibit appreciable tails as far out as 2.6--2.8 \AA.  Since the
nearest-neighbor C--Ba distance is only 3.1 \AA, the
partially filled Ba $5d$ orbitals overlap strongly with the
neighboring
C $2p$ states, which allows for
the formation of a covalent bond. This covalent bonding can be
further strengthened by
by allowing stronger overlap between the Ba and C states.
At self-consistency, this is
accomplished by reducing the ionicity of the Ba
atoms over what is expected
from the ionic model leading to longer ranged Ba d states. While
the covalent bonds are strengthened, the
Madelung stabilization
decreases and the actual degree of charge transfer is arrived at by
compromising between a purely covalent and purely ionic system.

To further characterize the degree of covalent vs. ionic bonding
character, we compare the valence-electron density in
${\rm Ba_6C_{60}}$ to that
of ${\rm K_6C_{60}}$.  In Fig.\ \ref{fig:density}(a) we show the
electron density
from the filled $t_{1u}$ band of ${\rm K_6C_{60}}$.
The radially directed
C-centered density lobes
are plainly evident, and very little contribution from K states
appears. (We also
note a slight but definite polarization of charge toward the K$^+$
ions). In Fig.\ \ref{fig:density}(b) we show the density from the
corresponding ($t_{1u}$)
band of ${\rm Ba_6C_{60}}$, and in Fig.\ \ref{fig:density}(c) the
$t_{1g}$
density. A substantial amount of Ba-derived density is evident in
each.
Nearest-neighbor C--Ba pairs
clearly show a substantial off-axis ``bent'' covalent bond,
accompanied by
strongly polarized back-bonding lobes.  Weaker bonds are formed
between next-nearest-neighbor C--Ba pairs. Given the
substantial covalent character accompanied by (incomplete) charge
transfer,
we conclude that bonding in
${\rm Ba_6C_{60}}$ is
best described as mixed covalent/ionic.

In summary, we have performed first-principles LDA calculations
on superconducting ${\rm Ba_6C_{60}}$ and find a number of unusual
features.  In contrast to the structurally identical compound
${\rm K_6C_{60}}$, for which we predicted full charge
transfer and ionic bonding in, for ${\rm Ba_6C_{60}}$
we find only partial charge transfer and mixed covalent/ionic bonding.
In contrast to the $A_3{\rm C}_{60}$ superconductors,
we calculate a zero-gap insulating band structure, which appears to be
at odds with the experimental finding of superconductivity.

We thank W.E. Pickett and E.J. Mele for many illuminating discussions.
This work was supported in part by the Laboratory for Research
on the Structure of Matter (University of Pennsylvania)
and in part by the Office of Naval Research.
%
%

%
%
\figure{Self-consistent electronic band structure for
${\rm Ba_6C_{60}}$.
The Fermi level is the energy zero. The dotted curves (for clarity
shown only  along the directions $\Gamma\!-\!N\!-\!H$) are the
relativistically
corrected bands described in the text.  \label{fig:bands}}
\figure{Total DOS near the Fermi level (top panel); atom- and
symmetry-projected partial DOS for carbon (middle panel)
and barium (bottom
panel).  For the partial DOS, $s$ components are shown as dotted curves
and are magnified by a factor of 10 (100) for C (Ba); there are only
negligible contributions from Ba $p$ states in this energy region.
Note also the factor of 3 scale difference between the C and Ba graphs.
\label{fig:pdos}}
\figure{Charge-difference function,  $\Delta Q(r)$, in a sphere
centered on a
Ba atom. The arrows denote the expectation values of $r$ for
atomic
Ba $5d$ and $6s$ states. \label{fig:deltaq}}
\figure{Valence-electron densities
for (a) the $t_{1u}$ band of
${\rm K_6C_{60}}$, (b) the $t_{1u}$ band of ${\rm Ba_6C_{60}}$,
and (c) the $t_{1g}$ band of ${\rm Ba_6C_{60}}$.
Each plotting plane contains four $C_{60}$
molecules (at the corners of the plot area) and four K$^+$ ions
(marked by large crosses). The projected C positions are marked by
small crosses, and the C$_{60}$ cage radius is shown by quarter
circles. A heavy solid line marks the nearest-neighbor C--Ba
internuclear axis. Adjacent contours
are separated by 0.0005 a.u. \label{fig:density}}
\begin{table}
\caption{Integrated muffin-tin (MT) charges, $Q(R)$, and charge
differences,
$\Delta Q(R)$, for the 4 inequivalent atoms in the ${\rm Ba_6C_{60}}$
cell.
C atoms are numbered by increasing distance from the nearest Ba atom.
Charge differences are defined with respect to overlapping
neutral atom electronic configurations.}
%
%
\begin{tabular}{lccc}
Atom & MT radius, $R$ (\AA)& $Q(R)$ & $\Delta Q(R)$ \\
\tableline
C$_1$ & \dec 0.70 &  \dec 4.05 & \dec $+$0.10 \\
C$_2$ & \dec 0.70 &  \dec 4.04 & \dec $+$0.10 \\
C$_2$ & \dec 0.70 &  \dec 4.06 & \dec $+$0.11 \\
Ba    & \dec 2.37 & \dec 59.24 & \dec $-$0.65 \\
\end{tabular}\label{table:charges}
\end{table}

\begin{references}
\bibitem{koretal92a}A. R. Kortan, N.
Kopylov, S. Glarum, E. M. Gyorgy, A. P. Ramirez,  R. M. Fleming, F.
A. Thiel and R. A.
Haddon, Nature (London) {\bf 355}, 529 (1992).
\bibitem{saiosh92}S. Saito and A. Oshiyama, Solid State Commun. {\bf
83}, 107 (1992).
\bibitem{koretal92b}A. R. Kortan {\it et al.}, preprint.
\bibitem{zhoetal91}O. Zhou, J. E. Fischer, N. Coustel, S. Kycia, Q.
Zhu, A. R.
McGhie,  W. J. Romanow, J. P. McCauley, A. B. Smith and D. E. Cox,
Nature {\bf
351}, 462 (1991).
\bibitem{erwped91}S. C. Erwin and M. R. Pederson, Phys. Rev. Lett.
{\bf 67}, 1610 (1991).
\bibitem{erwpic91}S. C. Erwin and W. E. Pickett, Science {\bf 254},
892 (1991).
\bibitem{erwpedpic90}S. C. Erwin, M. R. Pederson and W. E. Pickett,
Phys. Rev. B
{\bf 41},  10437 (1990).
\bibitem{erwpedunpub}S. C. Erwin and M. R. Pederson, unpublished.
\bibitem{richmondbb}M. R. Pederson, S. C. Erwin, K. A. Jackson and
L. L. Boyer, in {\it Physics and Chemistry of Finite Systems: From
Clusters to Crystals}, Vol.\ II, edited by P. Jena {\it et al.}
(Kuwar Academic Publishers, 1992), p.\ 1323.
\end{references}
\end{document}